\documentstyle[12pt,epsfig,amssymb,amstex]{article}
\begin{document}

\title{Characterizing self-organization and coevolution by ergodic invariants}
\author{R. Vilela Mendes \\
Grupo de F\'{\i }sica Matem\'{a}tica\\
Complexo Interdisciplinar, Universidade de Lisboa, \\
Av. Gama Pinto, 2 - P1699 Lisboa Codex, Portugal}
\date{}
\maketitle

\begin{abstract}
In addition to the emergent complexity of patterns that appears when many
agents come in interaction, it is also useful to characterize the dynamical
processes that lead to their self-organization. A set of ergodic invariants
is identified for this purpose, which is computed in several examples,
namely a Bernoulli network with either global or nearest-neighbor coupling,
a generalized Bak-Sneppen model and a continuous minority model.
\end{abstract}

\section{Introduction}

When a set of agents come in interaction there are, in general, thresholds
on the interaction strength above which the whole system self-organizes into
distinct patterns of collective behavior. If the dynamics of the agents and
their interactions are fixed for all time, the formation of collective
patterns is of no concern for the agents themselves. Rather, the
identification of patterns belongs to the information compression process of
the external observers. However, if the agents themselves are capable of
adaptation to the environment, then, the nature of the collective phenomena
strongly determines the process of coevolution.

Patterns of collective organization imply a correlation between the
configurations in phase space for the components of the system. Therefore
quantities like the entropy excess or the statistical complexity\cite
{Crutchfield1} \cite{Crutchfield2} \cite{Grassberger}, may be used to
characterize the complexity of the patterns that arise as a result of the
interactions. However, because the formation of the patterns is related to
the nature of the dynamical laws, it seems appropriate to look also for
quantities that relate directly to the dynamical process of
self-organization. Physically, the most relevant indices must be those that
are robust in a probability sense, that is, that are invariant almost
everywhere in the support of a physical measure. Therefore in this paper one
looks for ergodic invariants and, in particular, for those that emphasize
the dynamical relations between the system as a whole and its parts as well
as those that provide a dynamical characterization of the collective
structures.

Conditional exponents, corresponding to several splittings of the system,
when compared with the full Lyapunov exponents, are a measure of the
distinction between intrinsic dynamics and the dynamics that arises from the
interaction itself. On the other hand the conditional entropies
corresponding to cylindrical splittings of the phase space characterize the
relative independence of the parts in a collective system. The information
provided by the conditional entropies is, in general, not equivalent to the
information provided by the conditional exponents.

Another kind of information that is relevant to the characterization of
composite dynamical systems is the nature and origin of collective
structures. Structures, either temporal or spatial structures, are features
that occur at a (time- or space-) scale which is small as compared to the
scale of the individual components when in isolation. Temporal structures
are found to be related to the variation of the Lyapunov exponents with the
interaction strength. Of particular relevance is the critical behavior that
occurs in regions where some of the Lyapunov exponents approach zero. This
is used to define a structure index for temporal structures.

Conditional exponents, conditional exponent entropies and structure indices
are computed in several examples and related to their dynamical features.
They seem to relate well to the dynamical features in Bernoulli networks,
both with global and nearest-neighbor couplings, and in the generalized
Bak-Sneppen model studied in Sect.5.3. As for the last model that is
studied, a continuous version of the minority model, it belongs to a class
of models where the agents do not interact directly among themselves but
only through a common environment. In turn, the environment is a collective
variable that the agents themselves create. For this class of models,
conditional exponents, in particular, are not relevant and some other
invariants, beyond those developed in this paper, may be needed.

\section{Conditional exponents and conditional exponent entropies}

A dynamical system lives on the support of some measure $\mu $ which is left
invariant by the dynamics. An \textit{ergodic invariant} is a dynamical
characterization of this measure 
\[
I\left( \mu \right) =\lim_{T\rightarrow \infty }\frac{1}{T}\sum_{n=1}^{\textnormal{%
T}}\digamma \left( f^{n}x_{0}\right) \hspace{1in}x_{0}\textnormal{ }\mu -a.e. 
\]
Self-organization in a system concerns the dynamical relation of the whole
to its parts. The conditional Lyapunov exponents, introduced by Pecora and
Carrol\cite{Pecora1} \cite{Pecora2} in their study of synchronization of
chaotic systems, are quantities that in some sense try to separate the
intrinsic dynamics of each component from the influence of the other parts
in the system.

Let a mapping $f:M\rightarrow M$\ , with $M\subset R^{m}$\ and an invariant
measure $\mu $ define a $m-$dimensional dynamical system. The \textit{%
conditional exponents} associated to the splitting $\Sigma =R^{k}\times
R^{m-k}$\ are the eigenvalues $\xi _{i}^{(k)}$ and $\xi _{i}^{(m-k)}$ of the
limits 
\begin{eqnarray*}
&&\lim_{n\rightarrow \infty }\left( D_{k}f^{n*}(x)D_{k}f^{n}(x)\right) ^{%
\frac{1}{2n}} \\
&&\lim_{n\rightarrow \infty }\left( D_{m-k}f^{n*}(x)D_{k}f^{n}(x)\right) ^{%
\frac{1}{2n}}
\end{eqnarray*}
where $D_{k}f^{n}$\ and $D_{m-k}f^{n}$ are the $k\times k$\ and $m-k\times
m-k$ diagonal blocks of the full Jacobian. The conditional exponents are
good ergodic invariants.\medskip

\textit{Lemma}. Existence of the conditional exponents as well defined
ergodic invariants is guaranteed under the same conditions that establish
the existence of the Lyapunov exponents

Proof:

Let $\mu $\ be an ergodic $f-$invariant measure. Then, Oseledec's
multiplicative ergodic theorem, generalized for non-invertible $f$\ \cite
{Oseledec} \cite{Raghu} implies that if the map $T:M\rightarrow M_{m}$\ from 
$M$\ to the space of $m\times m$\ matrices is measurable and 
\begin{equation}
\int \mu (dx)\log ^{+}\left\| T(x)\right\| <\infty  \label{2.1}
\end{equation}
\ (with $\log ^{+}g=\max \left( 0,\log g\right) $) and if 
\begin{equation}
T_{x}^{n}=T(f^{n-1}x)\cdots T(fx)T(x)  \label{2.2}
\end{equation}
\ then 
\begin{equation}
\lim_{n\rightarrow \infty }\left( T_{x}^{n*}T_{x}^{n}\right) ^{\frac{1}{2n}%
}=\Lambda _{x}  \label{2.3}
\end{equation}
\ exists $\mu $\ almost everywhere.

If $T_{x}$\ is the full Jacobian $Df(x)$\ and if $Df(x)$\ satisfies the
integrability condition, the Lyapunov exponents exist $\mu -$a e.

If the Jacobian satisfies (\ref{2.1}), then the $m\times m$\ matrix formed
by the diagonal $k\times k$\ and $m-k\times m-k$\ diagonal blocks also
satisfies the same condition. Therefore conditional exponents too are
defined $\mu $-a. e..

Furthermore, under the same conditions, the set of regular points is Borel
of full measure and 
\[
\lim_{n\rightarrow \infty }\frac{1}{n}\log \left\| D_{k}f^{n}(x)u\right\|
=\xi _{i}^{(k)} 
\]
\ with $0\neq u\in E_{x}^{i}/E_{x}^{i+1}$\ , $E_{x}^{i}$\ being the subspace
of $R^{k}$\ spanned by eigenstates corresponding to eigenvalues $\leq \exp
(\xi _{i}^{(k)})$.\medskip

{\Large Conditional exponent entropies}

For an invariant measure $\mu $\ absolutely continuous with respect to the
Lebesgue measure of $M$\ or for measures that are smooth along unstable
directions (BRS\ measures), Pesin's identity\cite{Pesin} states that the sum
over positive Lyapunov exponents coincides with the Kolmogorov-Sinai
entropy. \ By analogy one defines the \textit{conditional exponent entropies}
associated to the splitting\ $R^{k}\times R^{m-k}$\ as 
\[
h_{k}(\mu )=\sum_{\xi _{i}^{(k)}>0}\xi _{i}^{(k)} 
\]
\ 
\[
h_{m-k}(\mu )=\sum_{\xi _{i}^{(m-k)}>0}\xi _{i}^{(m-k)} 
\]
These quantities, defined in terms of the conditional exponents, that are
good ergodic invariants, are also well-defined ergodic invariants. Here
these quantities are defined directly in terms of the conditional exponents.
In the next section we will describe an entropy construction in terms of the
dynamical refinements of partitions that correspond to the same splitting of
the phase space.\medskip

{\Large A measure of dynamical selforganization}

In information theory the mutual information $I(A:B)$\ is 
\[
I(A:B)=S(A)+S(B)-S(A+B) 
\]
\ By analogy one defines a \textit{measure of dynamical selforganization} $%
I(S,\Sigma ,\mu )$\ 
\[
I(S,\Sigma ,\mu )=\sum_{k=1}^{N}\left\{ h_{k}(\mu )+h_{m-k}(\mu )-h(\mu
)\right\} 
\]
\ The sum is over all relevant partitions $R^{k}\times R^{m-k}$ and $h(\mu )$
is the sum of the positive Lyapunov exponents

\[
h(\mu )=\sum_{\lambda _{i}>0}\lambda _{i} 
\]
$I(S,\Sigma ,\mu )$\ is also a well-defined ergodic invariant for the
measure $\mu $.

The Lyapunov exponents of a dynamical system measure the rate of information
production or, from an alternative point of view, they define the dynamical
freedom of the system, in the sense that they control the amount of change
that is needed today to have an effect on the future. In this sense the
larger a Lyapunov exponent is, the freer the system is in that particular
direction, because a very small change in the present state will induce a
large change in the future. From the point of view of the unit $k$ and of
the remaining subsystem, the quantity $h_{k}(\mu )+h_{m-k}(\mu )$ is
therefore the apparent dynamical freedom that they possess (or the apparent
rate of information production). The actual rate is in fact $h(\mu )$. Hence 
$I(S,\Sigma ,\mu )$ is a measure of the apparent excess dynamical freedom.

\section{Cylindrical partitions and conditional entropies}

Consider cylindrical partitions adapted to the splitting $R^{k}\times
R^{m-k} $, namely 
\[
\begin{array}{lll}
\eta ^{(k)} & = & \left\{ C_{1}^{(k)},C_{2}^{(k)},\cdots \right\} \\ 
\eta ^{(m-k)} & = & \left\{ C_{1}^{(m-k)},C_{2}^{(m-k)},\cdots \right\}
\end{array}
\]
where $C_{i}^{(k)}$ and $C_{i}^{(m-k)}$ are $k$ and $m-k$ -dimensional
cylinder sets in $R^{m}$.

Let now the $\zeta $ be a generator partition for the dynamics $\left( f,\mu
\right) $ and define the \textit{conditional entropies }associated to the
splitting\textit{\ }$R^{k}\times R^{m-k}$ by 
\[
\begin{array}{lll}
\frak{h}^{(k)} & = & \underset{\eta ^{(k)}}{\sup }\textnormal{ }\underset{%
n\rightarrow \infty }{\lim }\frac{1}{n+1}H\left( \zeta \vee f^{-1}\zeta \vee
\cdots \vee f^{-n}\zeta \mid \eta ^{(k)}\right) \\ 
\frak{h}^{(m-k)} & = & \underset{\eta ^{(m-k)}}{\sup }\textnormal{ }\underset{%
n\rightarrow \infty }{\lim }\frac{1}{n+1}H\left( \zeta \vee f^{-1}\zeta \vee
\cdots \vee f^{-n}\zeta \mid \eta ^{(m-k)}\right)
\end{array}
\]
$H\left( \chi \mid \eta \right) $ being 
\[
H\left( \chi \mid \eta \right) =-\underset{M/\eta }{\int }\sum_{i}\mu
\left( C_{i}^{(\chi )}\mid \eta \right) \ln \mu \left( C_{i}^{(\chi )}\mid
\eta \right) d\mu 
\]
That is, the conditional entropies are the supremum over all cylinder
partitions of the sum of the conditional Kolmogorov-Sinai entropies.

The conditional entropies have, in general, a meaning different from the
conditional exponent entropies defined before. They might nevertheless be
useful for the characterization of relative independence between the
components of a complex system. In the following, only the conditional
exponent entropies will be used.

\section{The structure index}

The Lyapunov exponents, as opposed to the conditional exponents, are a
global characterization of the dynamics. However they may also be used to
extract information, on the relation between the whole system and its parts.
If the dynamics of a single isolated unit is known, comparing this with the
spectrum of Lyapunov exponents of the coupled system, information may be
obtained on how the collective motion and coherent structures are organized.

A coherent structure (in a collective system) is a phenomenon that operates
at a scale very different from the scale of the component units in the
system. A structure in space is a feature at a length scale larger than the
characteristic size of the components and a structure in time is a
phenomenon with a time scale larger than the cycle time of the individual
components. A (temporal)\textit{\ structure index} may then be defined by

\begin{equation}
S=\frac{1}{N}\sum_{i=1}^{N_{s}}\frac{T_{i}-T}{T}  \label{4.1}
\end{equation}
where $N$ is the total number of components (degrees of freedom) of the
coupled system, $N_{s}$ is the number of structures, $T_{i}$ is the
characteristic time of structure $i$ and $T$ is the cycle time of the
isolated components (or, alternatively the characteristic time of the
fastest structure). A similar definition applies for a \textit{spatial
structure index}, by replacing characteristic times by characteristic
lengths.

Structures are collective motions of the system. Therefore their
characteristic times are the characteristic times of the separation
dynamics, that is, the inverse of the positive Lyapunov exponents. Hence,
for the temporal structure index, one may write 
\begin{equation}
S=\frac{1}{N}\sum_{i=1}^{N_{+}}\left( \frac{\lambda _{0}}{\lambda _{i}}%
-1\right)  \label{4.2}
\end{equation}
the sum being over the positive Lyapunov exponents $\lambda _{i}$. $\lambda
_{0}$ is the largest Lyapunov exponent of an isolated component or some
other reference value.

The temporal structure index diverges whenever a Lyapunov exponent
approaches zero from above. Therefore the index diverges at the points where
(in the separation dynamics) long time correlations develop.

\section{Examples}

\subsection{A globally coupled Bernoulli network}

The dynamical law is

\begin{equation}
x_{i}(t+1)=(1-c)f(x_{i}(t))+\sum_{j\neq i}\frac{c}{N-1}f(x_{j}(t))
\label{5.1}
\end{equation}
\ with $f(x)=2x$\ (mod. $1$).

The Lyapunov exponents are 
\begin{equation}
\begin{array}{lllllll}
\lambda _{1} & = & \log 2 &  &  &  &  \\ 
\lambda _{i} & = & \log \left( 2\left( 1-\frac{N}{N-1}c\right) \right) &  & 
\textnormal{with multiplicity} &  & N-1
\end{array}
\label{5.1a}
\end{equation}

Therefore, 
\begin{equation}
\begin{array}{cclcc}
h(\mu ) & = & \log 2+(N-1)\log \left( 2-\frac{2N}{N-1}c\right) & \textnormal{for}
& c\leq \frac{N-1}{2N} \\ 
& = & \log 2 & \textnormal{for} & c\geq \frac{N-1}{2N}
\end{array}
\label{5.2}
\end{equation}

The conditional exponents associated to the splitting $R^{1}\times R^{N-1}$\
are 
\begin{equation}
\xi ^{(1)}=\log (2-2c)  \label{5.3}
\end{equation}
\ and 
\begin{eqnarray}
&& 
\begin{array}{cccc}
\xi _{1}^{(N-1)} & = & \log \left( 2-\frac{2}{N-1}c\right) & \textnormal{once ;}
\end{array}
\label{5.4} \\
&& 
\begin{array}{ccccc}
\xi _{i}^{(N-1)} & = & \log \left( 2-\frac{2N}{N-1}c\right) & \textnormal{with
multiplicity} & N-2
\end{array}
\end{eqnarray}
\ Therefore, 
\begin{equation}
I(S,\Sigma ,\mu )=N\left( \log \left( 1-\frac{c}{N-1}\right) +\max \left(
\log (2-2c),0\right) -\max \left( \log \left( 2-\frac{2Nc}{N-1}\right)
,0\right) \right)  \label{5.5}
\end{equation}
\ which, in the limit of large $N$, becomes\ 
\begin{equation}
\begin{array}{ccccc}
I(S,\Sigma ,\mu ) & = & \frac{c^{2}}{1-c} &  & c\leq \frac{N-1}{2N} \\ 
& = & -c &  & c\geq \frac{1}{2}
\end{array}
\label{5.6}
\end{equation}
The variation of $I(S,\Sigma ,\mu )$ with the coupling intensity $c$ is
plotted in Fig.1. It grows until the synchronization point and then it
becomes negative. The transition is discontinuous in the large $N$
limit.\medskip

{\Large Structures}

Except for $c=c_{s}=\frac{1}{2}\frac{N-1}{N}$ the globally coupled Bernoulli
system is uniformly hyperbolic. For coupling strength $c<c_{s}$ the Lyapunov
dimension is $N$ and one expects a $BRS$- invariant measure absolutely
continuous with respect to the Lebesgue measure in $R^{N}$. The distribution
of the values taken by each unit $x_{i}$ is essentially flat and, for large $%
N$, the mean field seen by one unit has very small fluctuations.

However, as one approaches $c=c_{s}$ from below, one sees the dynamics
organizing itself into synchronized patches, with each patch maintaining an
approximately constant phase relation with the other patches.
Synchronization and phase locking effects are however not absolutely stable
phenomena. In the Figs.2 and 3 one shows the statistics of $\left|
x_{i}-x_{k}\right| $ and $x_{i}+x_{i+1}-2x_{i+2}$ . For $c=0.45$ one sees
clearly the phenomenon of clustering and synchronization with positive
Lyapunov exponents discussed before\cite{Vilela1}.

For $c<c_{s}$ all Lyapunov exponents are positive. However, near $c_{s}$
only one of the Lyapunov exponents is large whereas all the others are
nearly zero. That is, there is a fast \textit{separation dynamics }%
(sensitive dependence to initial conditions) in one direction and very slow
separation dynamics in all directions transversal to the fast one. The fast
separation direction corresponds to the eigenvector $(1,1,1,1,...,1)$

The slow separation dynamics in the transversal directions corresponds to
long wavelength effects in phase space which are the most sensitive to
boundary conditions and the available phase-space. Then, the slow temporal
structures beget non-uniform probability distributions in the linear
combinations of the variables that correspond to the slow eigenvalues. In
particular, $x_{i}-x_{i+1}$ corresponds to the eigenvector $%
(0,...,1,-1,0,...,0)$ and $x_{i}+x_{i+1}-x_{i+2}$ to $(0,...,1,1,-2,0,...,0)$%
.

The existence of structures near the transition points where one or more
Lyapunov exponents approach zero from above is an universal phenomena,
whereas the detailed form of the structures depends on the particular nature
of the available phase-space.

From (\ref{4.2}) and (\ref{5.1a}) one obtains for the structure index 
\[
\begin{array}{lllll}
S & = & \frac{N-1}{N}\left( \frac{\log 2}{\log 2\left( 1-\frac{N}{N-1}%
c\right) }-1\right) & \textnormal{for} & \frac{N}{N-1}c<0.5 \\ 
S & = & 0 & \textnormal{for} & \frac{N}{N-1}c>0.5
\end{array}
\]
For $\frac{N}{N-1}c>0.5$ the structure index vanishes because the
synchronized motion is effectively one-dimensional and the characteristic
time of the synchronized motion coincides with the characteristic time of
the individual units. The structure index is zero both for the uncoupled
case and the fully synchronized one and diverges at the synchronization
transition (Fig.2).

\subsection{Nearest neighbor coupling}

Let now

\[
x_{i}(t+1)=(1-c)f(x_{i}(t))+\frac{c}{2}\left(
f(x_{i+1}(t))+f(x_{i-1}(t))\right) 
\]
with $f(x)=2x$\ (mod. $1$)

The Lyapunov exponents are 
\[
\lambda _{k}=\log \left\{ 2\left( 1-c\right) +2c\cos \left( \frac{2\pi }{n}%
k\right) \right\} 
\]
$k=0,\cdots ,n-1$

For the conditional exponents, one is 
\[
\xi ^{(1)}=\log \left\{ 2\left( 1-c\right) \right\} 
\]
and the others are the logarithm of the eigenvalues of the matrix 
\[
\left( 
\begin{array}{ccccc}
2(1-c) & c & 0 & \cdots & 0 \\ 
c & 2(1-c) & c & \cdots & 0 \\ 
0 & c & 2(1-c) & c & \cdots \\ 
\cdots & \cdots & \cdots & \cdots & \cdots \\ 
0 & 0 & 0 & c & 2(1-c)
\end{array}
\right) 
\]
Fig.4 displays the measure of dynamical selforganization $I$ and the
structure index $S$ for this example with $N=500$. The dynamical behavior in
the nearest-neighbor coupled network is more complex than in the globally
coupled one, showing a greater diversity of distinct dynamical features.
This is reflected in the behavior of the ergodic invariants. For the
structure index, in particular, one notices the existence, above $c=0.5$, of
many points where it diverges. These points correspond to the crossing
through zero of each individual Lyapunov exponent. Also, the invariant $I$
displays three distinct regions, rather than two as in the globally coupled
network.

\subsection{A generalized Bak-Sneppen model}

As suggested by the examples above, the most interesting events (creation of
a large number of structures, for example) occur for particular values of
the interaction. Therefore, it would be desirable to have a way to adjust
the interaction strength in coevolution models. This is lacking in the
original Bak-Sneppen (B-S) model\cite{Bak}. Also, to define ergodic
invariants it is more convenient to have a deterministic dynamics. Notice
however that, even in stochastic models, it is possible to define a
parameter similar to a Lyapunov exponent, by analyzing the \textit{spread of
damage}\cite{Tamarit} \cite{Valleriani}.

The model studied in this section is defined by:

- $N$ species, each one assigned to a lattice point in a one-dimensional
lattice, each lattice point standing for an ecological niche. To each
lattice point $i$ one assigns a variable $x_{i}$ (with values between $0$
and $1$)

At each time step the site with the lowest $x_{i}$ and its two nearest
neighbors are chosen anew according to the law 
\[
\begin{array}{lll}
x_{i}(t+1) & = & \left( 1-c\right) f\left( x_{i}(t)\right) +\frac{c}{2}%
\left( f\left( x_{i+1}(t)\right) +f\left( x_{i-1}(t)\right) \right) \\ 
x_{i\pm 1}(t+1) & = & \left( 1-c\right) f\left( x_{i\pm 1}(t)\right) +\frac{c%
}{2}\left( f\left( x_{i}(t)\right) +f\left( x_{i\mp 1}(t)\right) \right)
\end{array}
\]
with $f(x)=2x$\ (mod. $1$)

When $c=0$ the function $f(x)$ is a pseudo-random number generator and the
model is equivalent to Bak-Sneppen's coarse grained model for evolution.
There are however some essential differences:

- B-S is a stochastic model whereas this one is deterministic. The updating
function being differentiable, the tools of ergodic theory are applicable
and Lyapunov exponents, conditional exponents and entropies may be computed
and used to characterize the dynamics.

- The interaction strength between neighboring species may be changed\medskip

\textbf{Parallel versus sequential dynamics}

If instead of sequential dynamics (selected by the smallest $x_{i}$) one had
chosen parallel dynamics, this model would be similar to the
nearest-neighbor coupled Bernoulli chain studied before. Then, a large
variety of qualitatively different dynamical behaviors are observed, which
result from the critical points at different values of $c$ where the
Lyapunov exponents cross zero.

When, instead of parallel dynamics, one imposes on the model a sequential
dynamics selected by the smallest $x_{i}$, what one is really doing is to
introduce a feature that simulates friction or resistance to change in the
dynamical system. This has some dramatic consequences for the computation of
the exponents in the limit of large $N$ . Instead of a tangent map matrix of
the type described before one has now the product of matrices which have
ones on the diagonal almost everywhere and only one non-trivial $3\times 3$
block. Therefore the Lyapunov exponents become, on the average 
\[
\begin{array}{llllll}
\lambda & \eqsim & \log \left( 2\right) ^{\frac{3}{N}} &  & \frac{N}{3} & 
\textnormal{times} \\ 
\lambda ^{^{\prime }} & \eqsim & \log \left( 2\left( 1-\frac{3}{2}c\right)
\right) ^{\frac{3}{N}} &  & \frac{2N}{3} & \textnormal{times}
\end{array}
\]
Therefore, for large $N$, all the Lyapunov exponents approach zero
independently of any other dynamical characteristics and the system is near
criticality. Therefore the fact that this type of system appears poised on
the edge of criticality is a consequence of the type of sequential dynamics
that is chosen.

For the structure index, in addition to the divergence effect obtained in
the limit of large $N$, one has other features that arise from the
interaction between the species, namely 
\[
\begin{array}{llllll}
c<\frac{1}{3} &  &  & S & = & \frac{\frac{N}{3\log 2}-\tau _{0}}{3\tau _{0}}%
+2\frac{\frac{N}{3\log 2\left( 1-\frac{3}{2}c\right) }-\tau _{0}}{3\tau _{0}}
\\ 
c>\frac{1}{3} &  &  & S & = & \frac{\frac{N}{3\log 2}-\tau _{0}}{3\tau _{0}}
\end{array}
\]
where $\tau _{0}$ is a reference characteristic time (characteristic time of
the individual dynamics or characteristic time of the fastest structure).
For definiteness one chooses $\tau _{0}=\frac{1}{3}\log 2$.

One sees that, besides the overall critical behavior arising from the
sequential dynamics, there is additional critical behavior at $c=\frac{1}{3}$
arising from the interactions in the coupled block of three species.

The average conditional exponents are 
\[
\begin{array}{lll}
\mu ^{\left( 1\right) } & \eqsim & \log \left( 2\left( 1-c\right) \right) ^{%
\frac{3}{N}} \\ 
\mu ^{\left( N-1\right) } & \eqsim & 
\begin{array}{l}
\log \left( 2\left( 1-\frac{1}{2}c\right) \right) ^{\frac{3}{N}} \\ 
\log \left( 2\left( 1-\frac{3}{2}c\right) \right) ^{\frac{3}{N}}
\end{array}
\end{array}
\]

and the measure of dynamical self-organization is 
\[
\begin{array}{llllll}
c<\frac{1}{3} &  &  & I & = & 3\left( \log \left( 1-c\right) +\log \left( 1-%
\frac{c}{2}\right) -\log \left( 1-\frac{3}{2}c\right) \right) \\ 
\frac{1}{2}>c>\frac{1}{3} &  &  & I & = & 3\left( \log 2+\log \left(
1-c\right) -\log \left( 1-\frac{1}{2}c\right) \right) \\ 
1>c>\frac{1}{2} &  &  & I & = & 3\log \left( 1-\frac{c}{2}\right)
\end{array}
\]
The values of the structure index and the self-organization measure for the
generalized Bak-Sneppen model are plotted in Fig.5. 

Not only the ergodic
invariants, but also variables like the barrier size, show a marked
dependence on the coupling strength. Numerically computed barrier sizes are
shown in Fig.6. 

One sees that from $c=0$ to $c=\frac{1}{3}$ the barrier
stays close to the original B-S value. Then, after $c=\frac{1}{3}$ (the
additional critical point of the structure index) it grows to a plateau
above 0.95. Eventually, as $c$ increases further, the barrier size comes
back to the original B-S value. This is probably related to the fact that
for large values of the coupling, when neighboring species become
synchronized, there is an almost random evolution of three-species blocks.

The scaling of avalanche sizes is another useful characterization in models
of this type. The avalanche size is defined, as usual, by the number of time
steps below a fixed threshold. With thresholds for the avalanche definition
at $0.65$ in the first case and $0.95$ in the second, a comparison was made
of the $c=0$ and the $c=0.5$ cases. In both cases a power law is obtained, $%
N(s)\thicksim s^{-\alpha }$, with a $20\%$ larger exponent $\alpha $ in the $%
c=0.5$ case.

\subsection{A continuous minority model}

The minority model introduced by Challet and Zhang,\cite{Zhang}, inspired in
Brian Arthur's bar model\cite{Arthur}, as well as most market models,
belongs to a class of models in which the agents do not interact directly
among themselves, but only through a common environment. On the other hand,
the environment is a collective functional that the agents themselves
create. This kind of inter-agents interaction, through an external medium,
that they themselves collectively create, has some specific dynamical
consequences.

Here one studies a continuous version of the minority model. For each one of 
$N$ agents there is a variable $x_{i}$ with values in the interval $[0,1)$.
The average of the values of $x_{i}$ defines a ''mean field'' $\overline{x}%
(t)$ at time $t$%
\[
\overline{x}(t)=\frac{1}{N}\sum_{i}x_{i}(t) 
\]
The time evolution of each $x_{i}$ is only a function of $\overline{x}$ at
times $t,t-1,\cdots ,t-M$. Let $c\in [0,1)$ (called the \textit{cut}) define
a partition of the interval into $\left\{ A=[0,c),B=[c,1)\right\} $. At each
time $t$ if $x_{i}(t)$ is in one of these intervals and $\overline{x}(t)$ is
in the other, agent $i$ wins a point, otherwise it wins nothing. Hence one
has the dynamics 
\[
x_{i}(t+1)=f_{i}\left( \overline{x}(t),\cdots ,\overline{x}(t-M)\right) 
\]
and a payoff 
\[
m_{i}(t+1)=m_{i}(t)+\frac{1}{2}\left( 1-\textnormal{sign}\left\{ \left( \overline{x%
}(t)-c\right) \left( x_{i}(t)-c\right) \right\} \right) 
\]

Each agent has his own function $f_{i}$, called the \textit{strategy} of
agent $i$. The time-delayed law may be converted into a single map in a $%
(M+1)N$ space 
\[
\left( 
\begin{array}{l}
x_{i}(t) \\ 
x_{i}(t-1) \\ 
\vdots \\ 
x_{i}(t-M)
\end{array}
\right) \rightarrow \left( 
\begin{array}{l}
f_{i}\left( \overline{x}(t),\cdots ,\overline{x}(t-M)\right) \\ 
x_{i}(t) \\ 
\vdots \\ 
x_{i}(t-M+1)
\end{array}
\right) 
\]
It is easy to see that, for models of this kind, the conditional exponents
play a negligible role. Take for example the case $M=0$. Consider the
Jacobian 
\[
J_{p}=\frac{\partial x_{i}(t+p)}{\partial x_{j}(t)} 
\]
The eigenvalues of $J^{T}J$ are $N-1$ zeros and one 
\[
\mu =N\left( \frac{1}{N^{2}}\sum_{i}f_{i}^{^{\prime }2}(t+p)\right) \left( 
\frac{1}{N}\sum_{i}f_{i}^{^{\prime }}(t+p-1)\right) ^{2}\cdots \left( \frac{1%
}{N}\sum_{i}f_{i}^{^{\prime }}(t)\right) ^{2} 
\]
Taking the $p\rightarrow \infty $ limit in 
\[
\lim_{p\rightarrow \infty }\log \left( J^{T}J\right) ^{\frac{1}{2p}} 
\]
one obtains a single non-trivial Lyapunov exponent 
\[
\lambda =\lim_{p\rightarrow \infty }\frac{1}{p}\log \left\{ \left( \frac{1}{N%
}\sum_{i=1}^{N}f_{i}^{^{\prime }}(t+p-1)\right) \cdots \left( \frac{1}{N}%
\sum_{i=1}^{N}f_{i}^{^{\prime }}(t)\right) \right\} 
\]
For the conditional exponents 
\[
\begin{array}{lll}
\xi _{i}^{(1)} & = & \lim_{p\rightarrow \infty }\frac{1}{p}\log \left\{
\left( \frac{1}{N}f_{i}^{^{\prime }}(t+p)\right) \cdots \left( \frac{1}{N}%
f_{i}^{^{\prime }}(t)\right) \right\} \\ 
\xi _{i}^{(N-1)} & = & \lim_{p\rightarrow \infty }\frac{1}{p}\log \left\{
\left( \frac{1}{N}\sum_{j\neq i}f_{j}^{^{\prime }}(t+p-1)\right) \cdots
\left( \frac{1}{N}\sum_{j\neq i}f_{j}^{^{\prime }}(t)\right) \right\}
\end{array}
\]
Then, for a sufficiently large number of agents, the conditional exponent $%
\xi _{i}^{(1)}$ cannot be positive and the self-organization measure will
always vanish in the large $N$ limit. The situation does not change if a
non-zero memory size $M$ is considered. Then the conditional exponent $\xi
_{i}^{(1)}$ is computed by the product of blocks of the form 
\[
\left( 
\begin{array}{llll}
\frac{1}{N}f_{i}^{(1)} & \frac{1}{N}f_{i}^{(2)} & \cdots & \frac{1}{N}%
f_{i}^{(M+1)} \\ 
1 & 0 & \cdots & 0 \\ 
0 & 1 & \cdots & 0 \\ 
\cdots & \cdots & \cdots & \cdots \\ 
0 & 0 & \cdots & 0
\end{array}
\right) 
\]
where $f_{i}^{(k)}$ denotes the $k$-argument derivative of $f_{i}$ .

For large $N$ the quantities $\frac{1}{N}f_{i}^{(k)}$ are very small and
once again the conditional exponent $\xi _{i}^{(1)}$ cannot be positive. The
vanishing of the $I(S,\Sigma ,\mu )$ invariant is easy to understand from
the qualitative interpretation given in Sect.2. For models of this type, the
contribution of each individual agent to his own evolution is extremely
small and therefore the difference between $h_{k}(\mu )+h_{N-k}(\mu )$ and $%
h(\mu )$ must be negligible.

In the original (discrete) minority model each agent is equipped with
several strategies choosing, at each time, the one with the best virtual
record. Here each agent has a single strategy which however may be changed
according to the following scheme:

- At the start all strategies are chosen at random from a pool of functions;

- After each 10 time steps, the 10 worst performing strategies (in the last
10 steps) are selected for replacement. The three worst ones are replaced by
three new strategies chosen at random from the function pool and the
remaining seven are replaced by strategies that copy the seven best ones
with a small random error.

For definiteness consider the functions to be linear regressions with
coefficients $\alpha _{k}^{(i)}$ taken at random from the interval $[-K,K]$. 
\[
x_{i}(t+1)=\sum_{k=0}^{M}\alpha _{k}^{(i)}\overline{x}(t-k) 
\]

For sufficiently large $K$ this system self-organizes into configurations
away from random choice. The most relevant parameter to track the system
behavior is 
\[
P(t)=\sum_{i}m_{i}(t)
\]
the total payoff at time $t$. In the following table one compares the
average and standard deviations of $P$ for random choice of the $x_{i}$'s in
the interval $[0,1)$ with those obtained from numerical simulations of the
model for several values of $K$ and cut $c$ and $M=2$. 
\[
\begin{tabular}{||l||l||l||l||l||l||l||}
\hline\hline
Cut & $\overline{P}_{\textnormal{rand}}$ & $\sigma (P)_{\textnormal{rand}}$ & $K$ & $%
\overline{P}$ & $\sigma (P)$ & $\sum_{i}\lambda _{i}$ \\ \hline\hline
0.6 & 0.4 & 0.049 & 
\begin{tabular}{l}
1 \\ 
1.5 \\ 
2
\end{tabular}
& 
\begin{tabular}{l}
0.494 \\ 
0.49 \\ 
0.486
\end{tabular}
& 
\begin{tabular}{l}
0.1 \\ 
0.099 \\ 
0.096
\end{tabular}
&  \\ \hline\hline
0.7 & 0.3 & 0.0459 & 
\begin{tabular}{l}
0.5 \\ 
1 \\ 
1.5 \\ 
2 \\ 
4 \\ 
5
\end{tabular}
& 
\begin{tabular}{l}
0.495 \\ 
0.499 \\ 
0.496 \\ 
0.493 \\ 
0.483 \\ 
0.481
\end{tabular}
& 
\begin{tabular}{l}
0.186 \\ 
0.17 \\ 
0.165 \\ 
0.157 \\ 
0.15 \\ 
0.157
\end{tabular}
& 
\begin{tabular}{l}
-4.38 \\ 
\\ 
-1.13 \\ 
-1.05 \\ 
\\ 
-2.2
\end{tabular}
\\ \hline\hline
\end{tabular}
\]
$\overline{P}$ and $\sigma (P)$ are the average and standard deviations of $%
P(t)$ and $\overline{P}_{\textnormal{rand}}$ and $\sigma (P)_{\textnormal{rand}}$ the
values that are obtained for random choice of the $x_{i}$'s in the interval $%
[0,1)$.

One sees that the average value of $P$ is systematically larger than the
random value, but with a much larger standard deviation. Of particular
significance is the actual probability distribution of this global variable.
In Figs.7 and 8 one shows this distribution for $c=0.6$ and $c=0.7$. Instead of a
Gaussian distribution with a small standard deviation, that would be
obtained in the random case, the model organizes itself to have a
asymmetrical distribution, with an increased average value, which for large
cuts splits into two peaks

The dynamics of the total payoff $P(t)$%
\[
P(t)=\sum_{i}\frac{1}{2}\left( 1-\textnormal{sign}\left\{ \left( \overline{x}%
(t)-c\right) \left( \textnormal{mod}\left( \sum_{k=1}^{M+1}\alpha _{k}^{(i)}%
\overline{x}(t-k),1\right) -c\right) \right\} \right) 
\]
depends only on the dynamics of the average values 
\[
\left( 
\begin{array}{l}
\overline{x}(t) \\ 
\overline{x}(t-1) \\ 
\vdots \\ 
\overline{x}(t-M)
\end{array}
\right) \rightarrow \left( 
\begin{array}{l}
\sum_{k=0}^{M}\frac{1}{N}\sum_{i}\alpha _{k}^{(i)}\overline{x}(t-k)+\frac{1}{%
N}\sum_{i}\theta _{i}(t) \\ 
\overline{x}(t) \\ 
\vdots \\ 
\overline{x}(t-M+1)
\end{array}
\right) 
\]
where the integers $\theta _{i}(t)\in Z$ originate from the $\textnormal{mod}.1$
operation in the dynamics of the $x_{i}$'s. It is the dynamics of the
average values that determines the global behavior of the system. The
Lyapunov exponents of this dynamics is computed from the Jacobian matrices 
\[
\left( 
\begin{array}{llll}
\frac{1}{N}\sum_{i}\alpha _{0}^{(i)} & \frac{1}{N}\sum_{i}\alpha _{1}^{(i)}
& \cdots & \frac{1}{N}\sum_{i}\alpha _{M}^{(i)} \\ 
1 & 0 & \cdots & 0 \\ 
0 & 1 & \cdots & 0 \\ 
\cdots & \cdots & \cdots & \cdots \\ 
0 & 0 & \cdots & 0
\end{array}
\right) 
\]
Typically, in the (statistically-) stable state of the model, the Lyapunov
exponents are all negative. They were computed for typical configurations
obtained after many iterations for several values of $K$ at $c=0.7$. This is
the meaning of the last column in the table above. Although the Lyapunov
exponents are negative, the dynamics is non-trivial. Without the non-linear
effect of the $\theta _{i}$'s the system would stabilize in a fixed point.
Instead, for large $N$, it has very many periodic orbits and the evolution
mechanism, emphasizing those that have a higher payoff for each cut, acts a
selection mechanism that drives the system to a particular class of orbits.
In the Fig.9 one shows the typical time-behavior of the mean value $%
\overline{x}(t)$ after many iterations of the model. The fluctuations that
are observed reflect not only the high period of the orbits but also the
change of dynamics imposed by the selection mechanism that is operating all
the time.

Thus, the dynamics of this continuous minority model becomes well
understood. However the role of the ergodic invariants discussed in Sects.
2-4 is not so clear. On the one hand, as explained above, the ''mean-field''
dynamics of the model makes the conditional exponents trivial. On the other
hand the structure index invariant is suited mainly to detect structural
transitions when a parameter is varied. Here however one finds out that the
dynamical structure is largely independent of the parameters that have been
explored $(K$ and $c)$. The structure index might however become useful if
this model is embedded in some larger class of models.

Another variable that is relevant in the study of collective model of this
type is the survival time of each strategy, that is the time interval until
it follows in one of the (locally) worst strategies. From the simulations
that were performed an exponential behavior ($\exp (-\xi t)$) is obtained,
rather than a power law as in the avalanche sizes of the generalized
Bak-Sneppen model. Also, the exponent is nearly constant, $\xi \simeq 0.004$%
, in the range of parameters that was explored

\section{Remarks and conclusions}

A fascinating aspect of complex systems, and even more of complex adaptive
systems, is that the behavior of the whole is so different from and richer
than the behavior of the parts. For lack of a precise theory of collective
behavior, all kinds of new features that appear in the whole are called 
\textit{emergent properties}. From simple systems with simple rules, complex
macropatterns emerge. To understand why this is so and what universal
features, if any, underlie this phenomenon is a challenging task. It is also
of practical importance because emergence is ubiquitous in the universe
around us.

An almost general rule, in the emergence of macropatterns, is the formation,
through interactions, of subassemblies which combine with similar
subassemblies, with the structure at each level constraining what emerges at
the next level. The nature of the interactions between the agents, and then
between the subassemblies, is the key to the understanding of the
macropatterns. This is the reason why, to understand emergence and
self-organization, it is essential to characterize the interaction dynamics
and, in particular, the robust properties of this dynamics, that is, its
ergodic properties.

One dynamical effect, identified in this paper, which generates collective
patterns is the approach to zero from above of one of the Lyapunov exponents
of the global system. The question is how frequently should one expect this
situation to occur. Two basic mechanisms were identified:

- When there is a natural limitation on the range of values that the state
variable of each individual agent can take, the coupling must be convex, as
in the examples studied in this paper. Then, the convex coupling leads to an
overall contracting effect and Lyapunov transitions are expected when the
coupling increases. In spatially extended systems, for example, even if the
interaction law does not change, a change in density would imply an
effective coupling increase. Therefore in a evolving system where the number
of agents changes in time (but the available space remains fixed), effects
of the type described here might be expected to arise when the population
density changes.

That, at the transition regions between chaos and order, evolving systems
display interesting structural properties was suggested in the past by
several authors\cite{Langton} \cite{Kauffman}. Why some natural systems
might have evolved to such narrow regions in parameter space is, to a large
extent, an open question. The density-dependent increase of the effective
interaction and the contracting effect implied by the convex coupling, when
the amount of available phase-space remains constant, is a dynamical
mechanism that might explain, in some cases, the evolution towards the
transition regions.

- Another mechanism leading to Lyapunov exponents that are positive but
close to zero, occurs when agents with sensitive dependent dynamics interact
via a friction mechanism. This resistance to change has as a consequence
that only the agents under the largest stress will be allowed to evolve. For
a large number of agents, this sequential dynamics leads to an effective
Lyapunov exponent close to zero. This is a situation that seems to occur in
many examples of what has been called \textit{self-organized criticality}.
These systems appear poised on the edge of criticality as a consequence of
this type of sequential dynamics.

The approach to zero of the Lyapunov exponents corresponds to the divergent
points of the (temporal) \textit{structure index}. Another important
characterization of the collective system is obtained by invariants
constructed from the \textit{conditional exponents}. These are quantities
that distinguish the intrinsic dynamics of each agent from the influence of
the other parts in the system. In particular the \textit{measure of
dynamical selforganization}, discussed in Sect.2, characterizes the excess
dynamical freedom that would be perceived by the agents themselves.

Structure indices and the invariants constructed from the conditional
exponents and conditional entropies do not, however, exhaust the parameters
needed to characterize the dynamical and probabilistic behavior of
collective systems. Scaling exponents, for example, are currently used and,
as seen in the generalized Bak-Sneppen model, may be related to the other
invariants. A technique which might also be useful, in the future, is the
generalized spectral decomposition, in particular the relation between the
spectrum of the Koopman operator of individual agents with the spectrum of
the whole system.\bigskip

{\Large Figure captions}

Fig.1 - Structure index and self-organization measure for a globally coupled
Bernoulli network

Fig.2 - Distribution of the variable $\left| x_{i}-x_{j}\right| $ for
several values of the coupling and $N=100$

Fig.3 - Distribution of the variable $x_{i}+x_{i+1}-2x_{i+2}$ for several
values of the coupling and $N=100$

Fig.4 - Structure index and self-organization measure for a nearest-neighbor
coupled Bernoulli network

Fig.5 - Structure index and self-organization measure for the generalized
Bak-Sneppen model

Fig.6 - Barrier size for several values of the coupling in the generalized
Bak-Sneppen model

Fig.7 - Distribution of $P$ for $c=0.6$

Fig.8 - Distribution of $P$ for $c=0.7$

Fig.9 - Typical time behavior of the mean value in the continuous minority
model

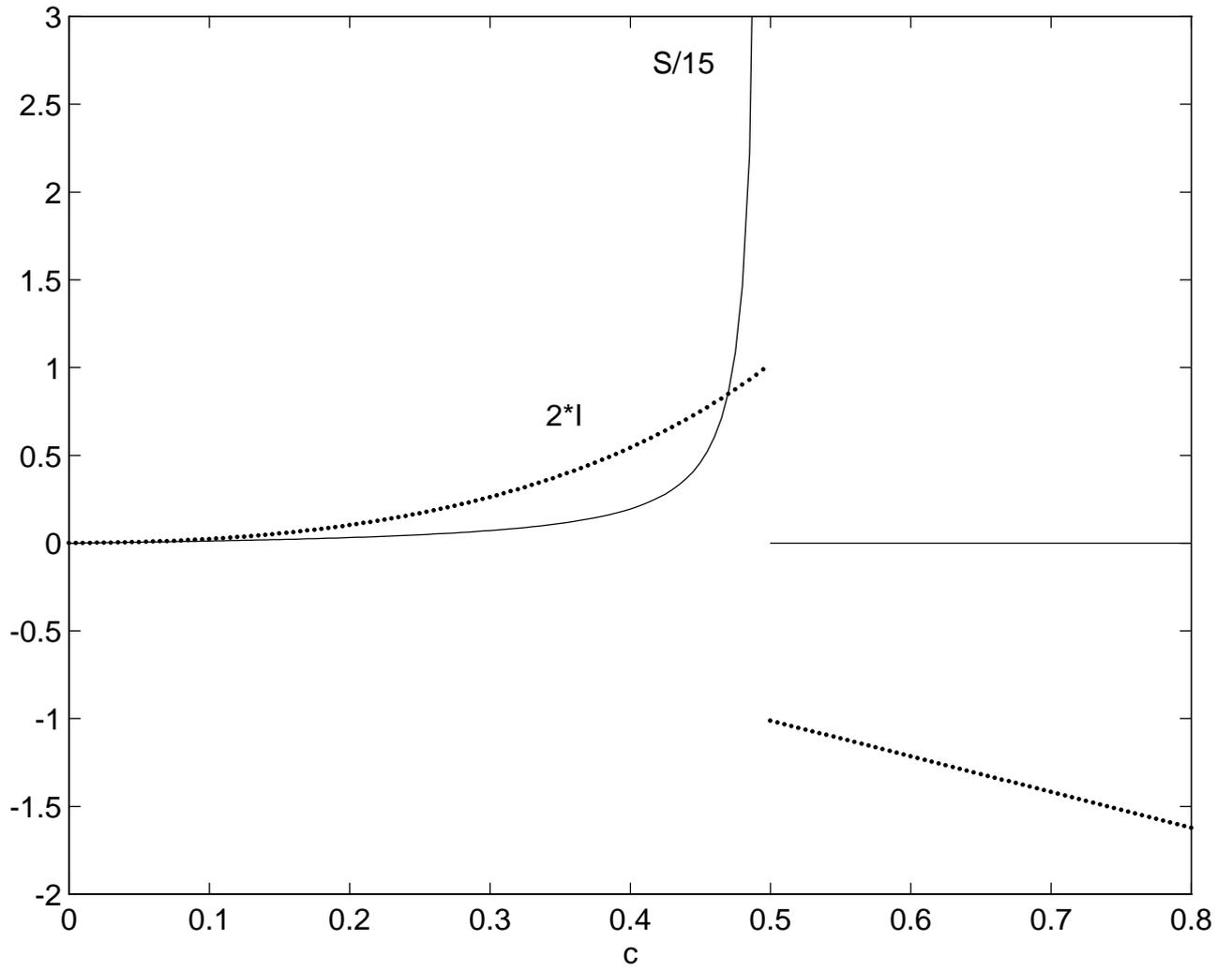
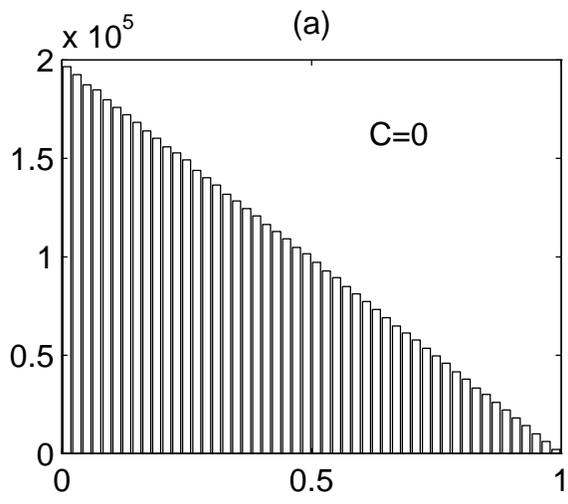
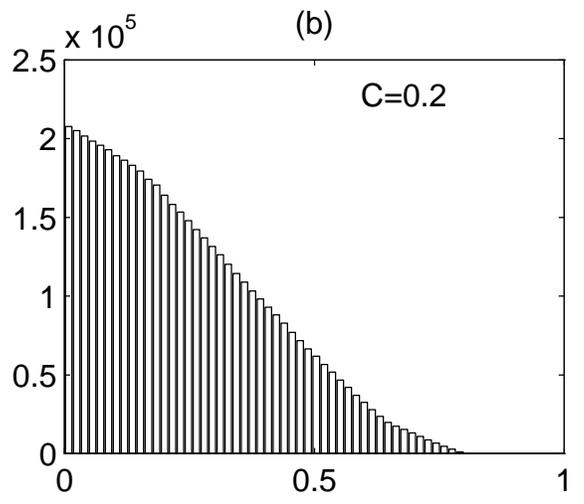
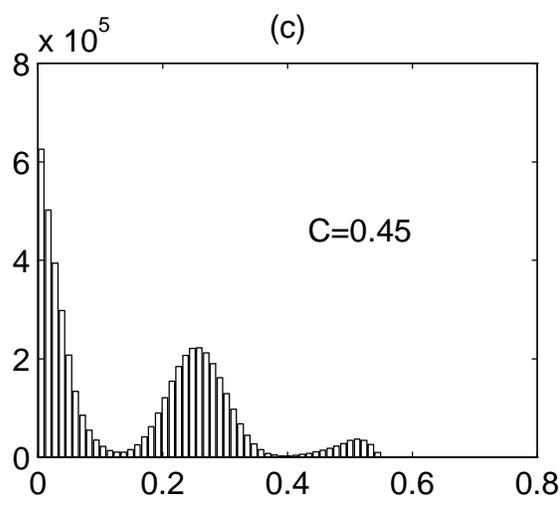
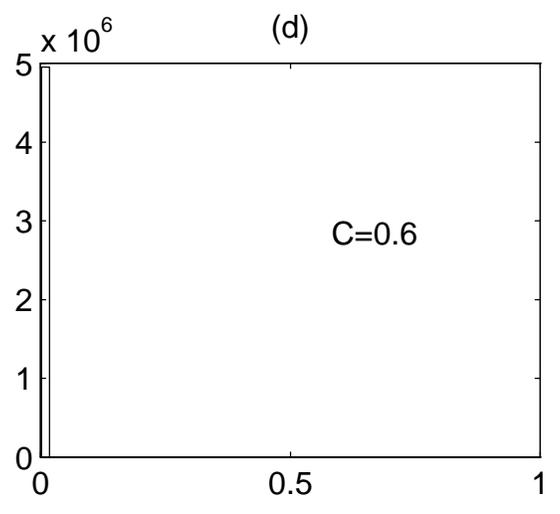
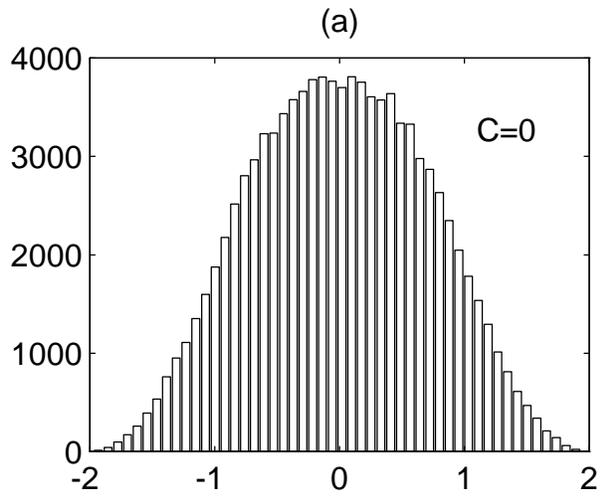
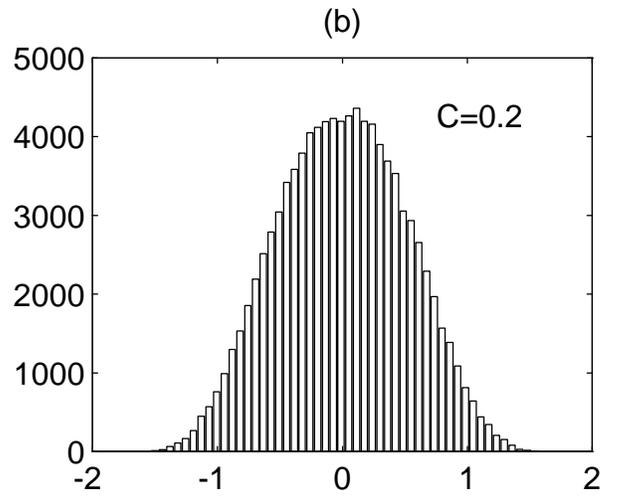
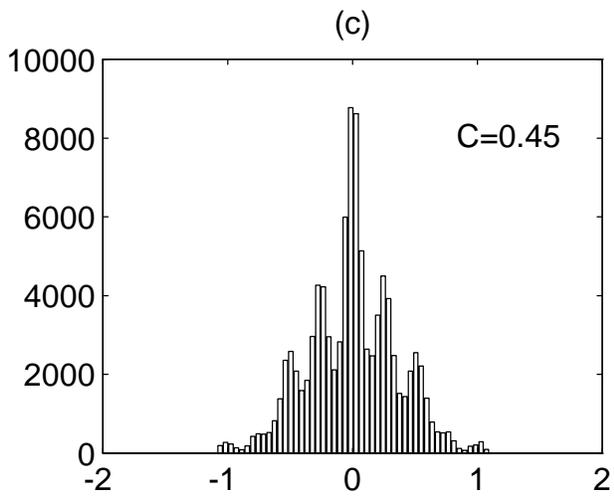
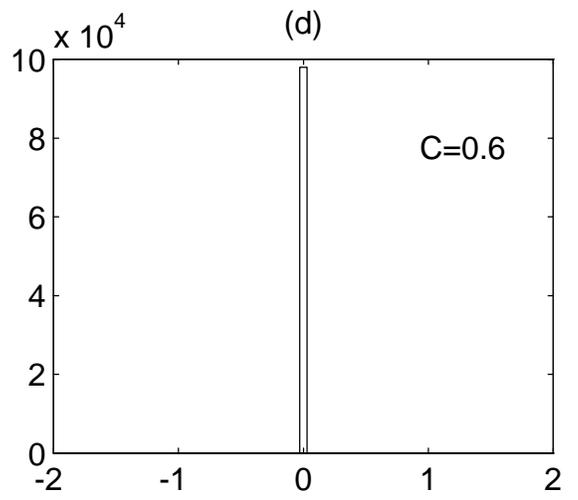
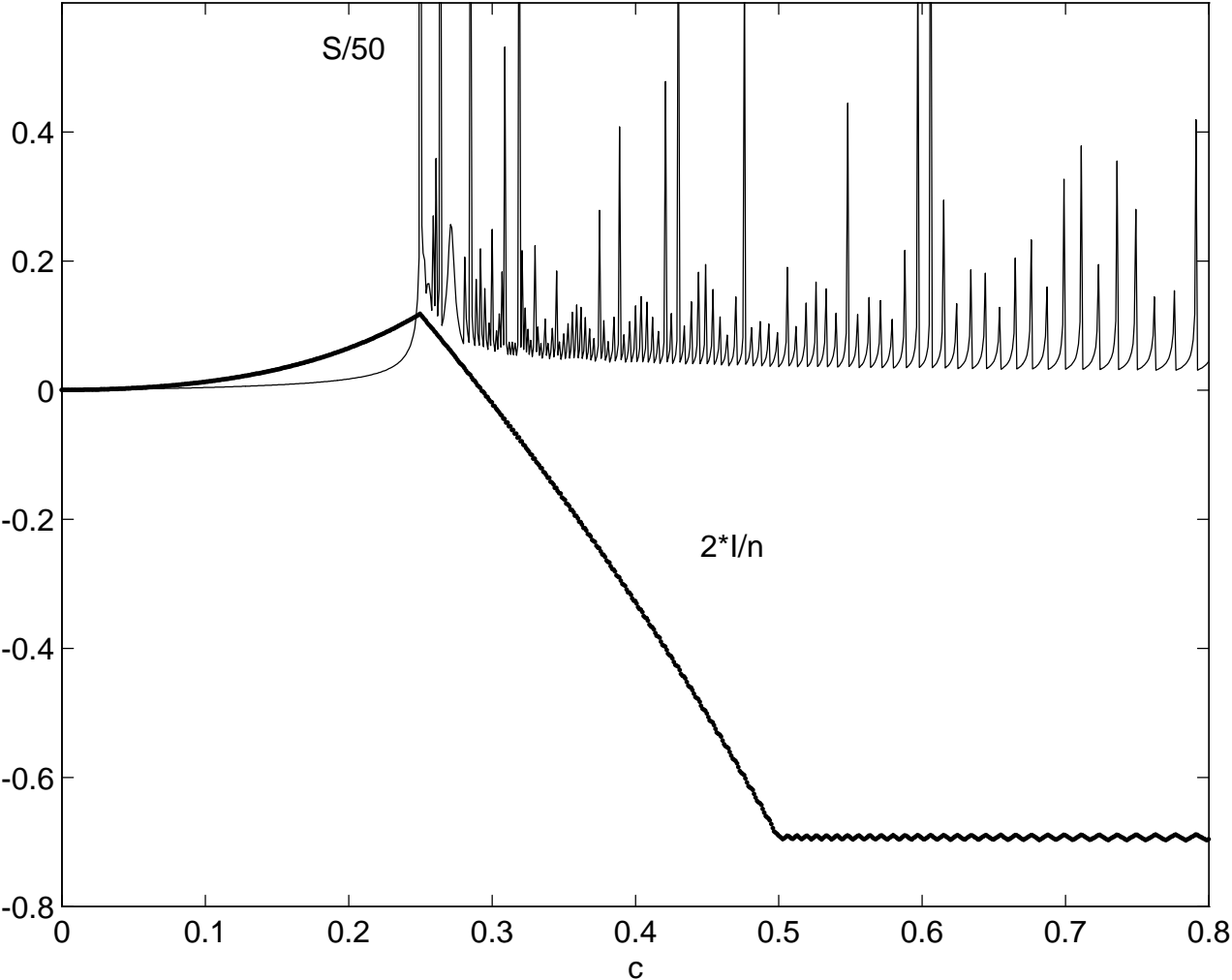
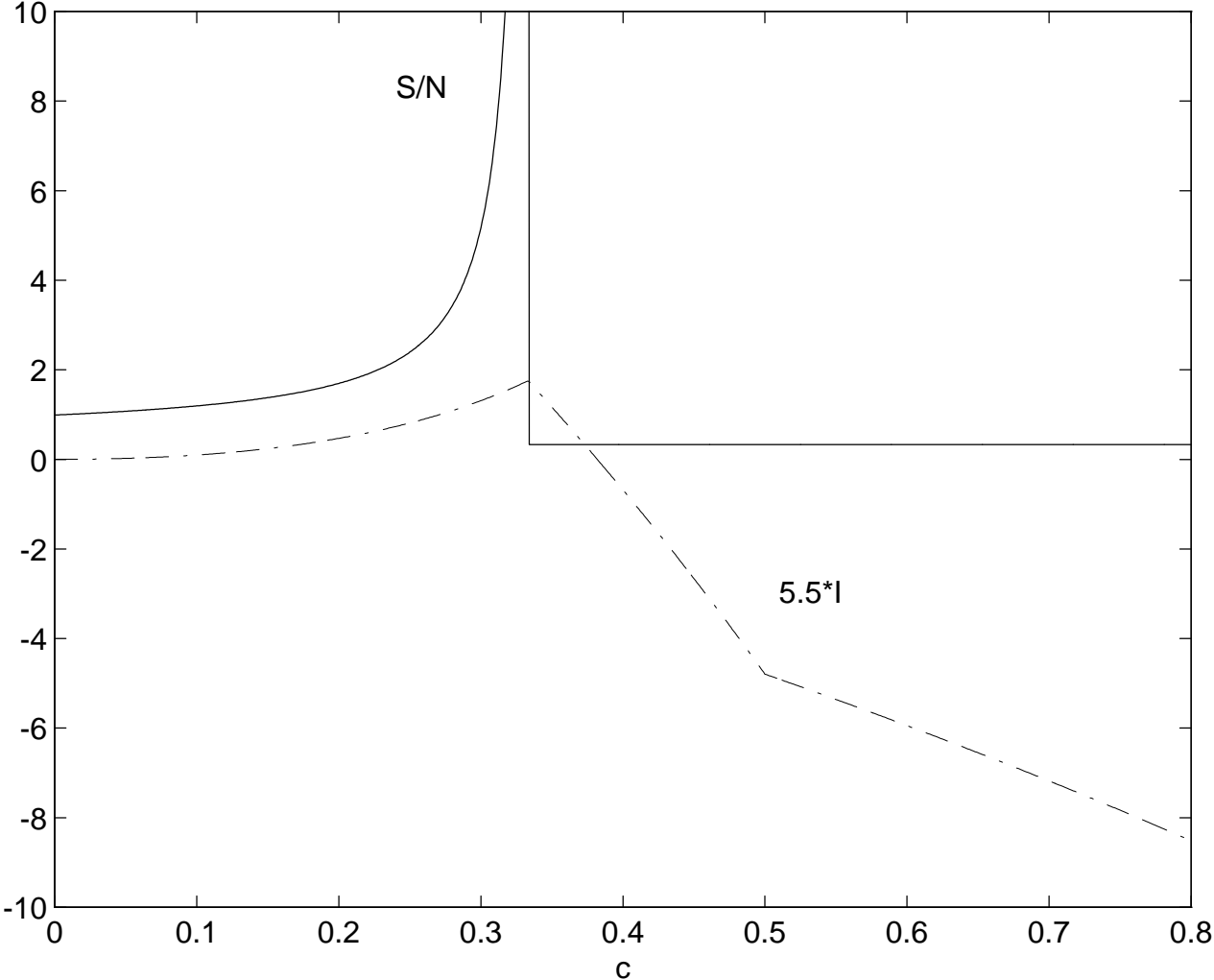
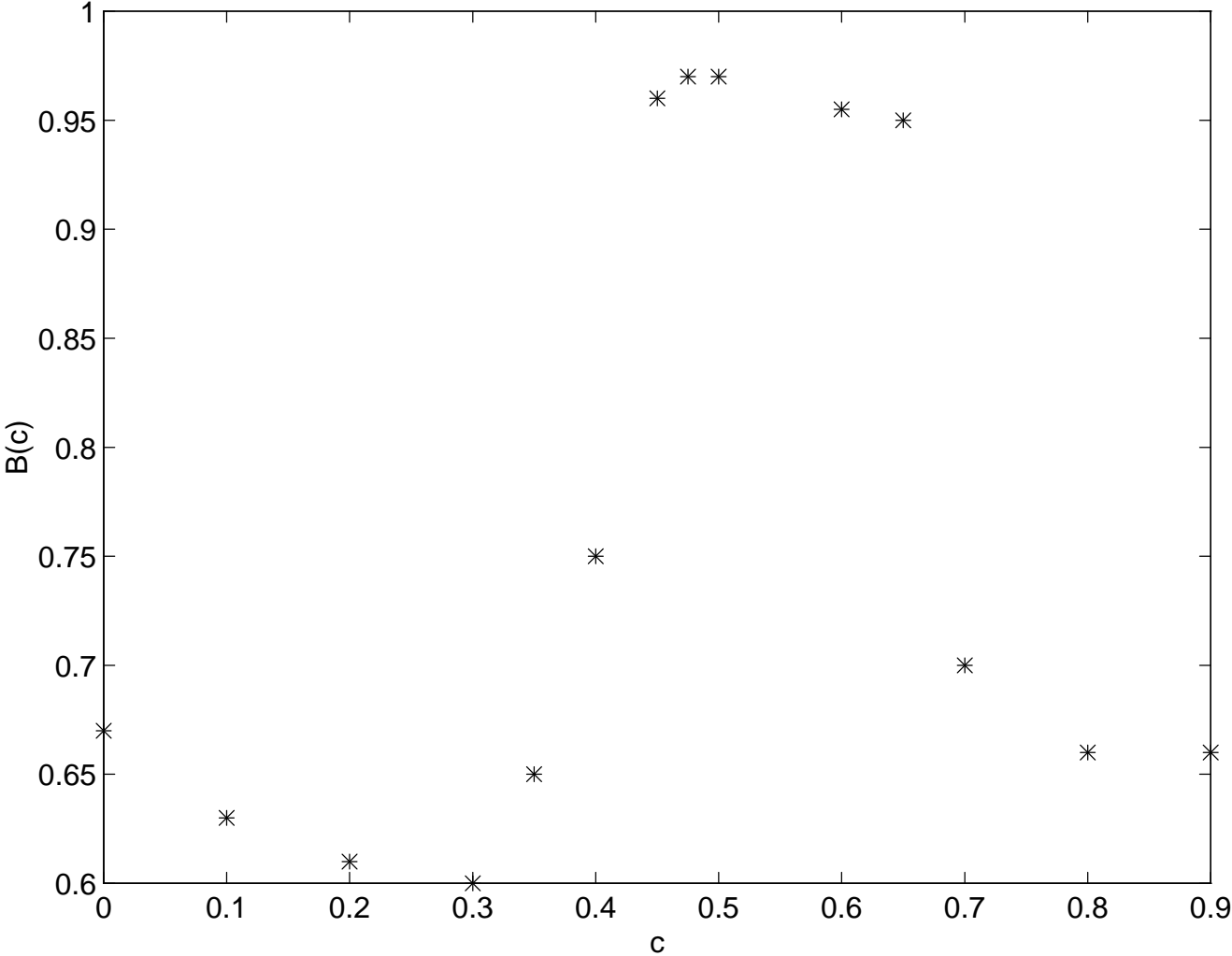
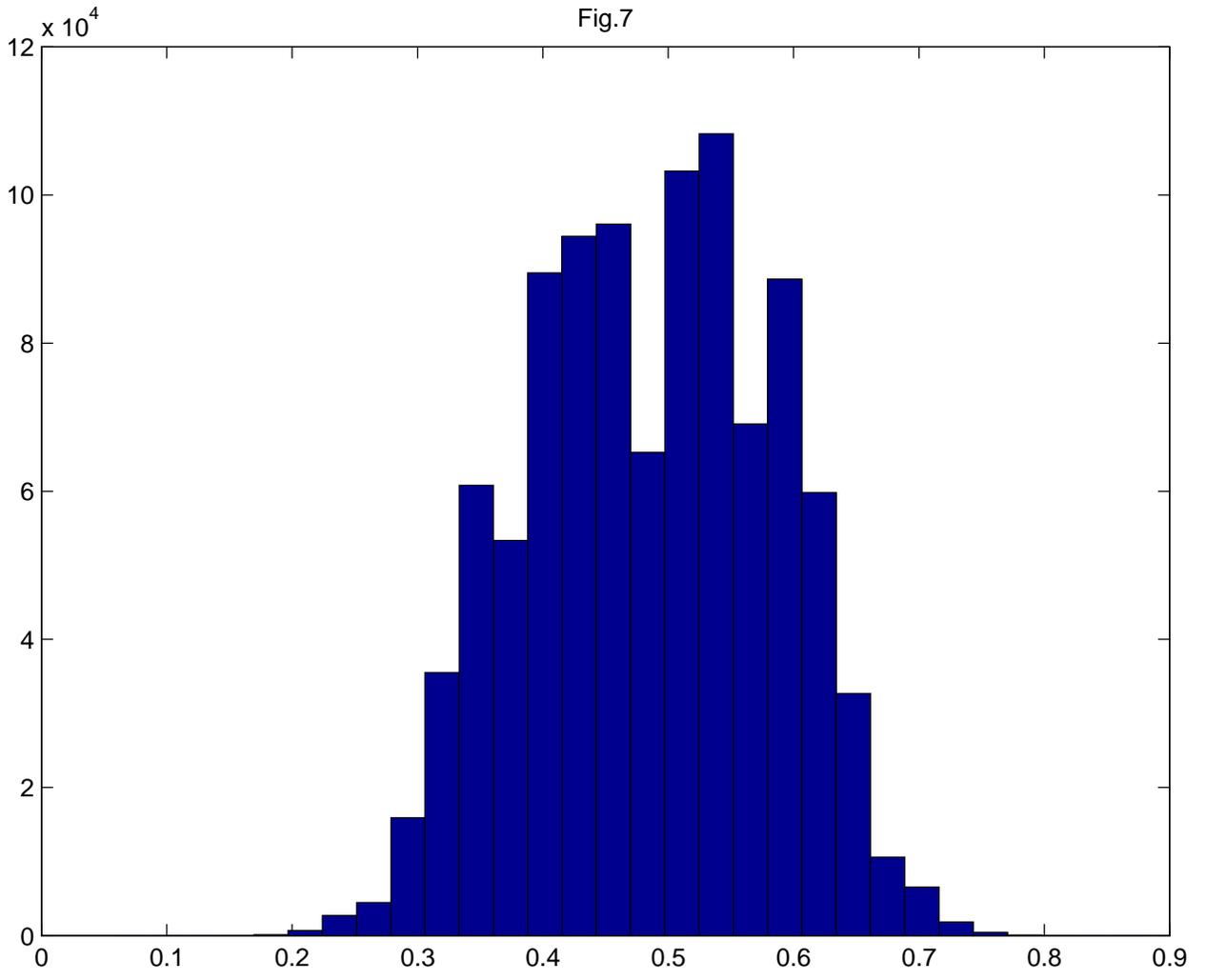
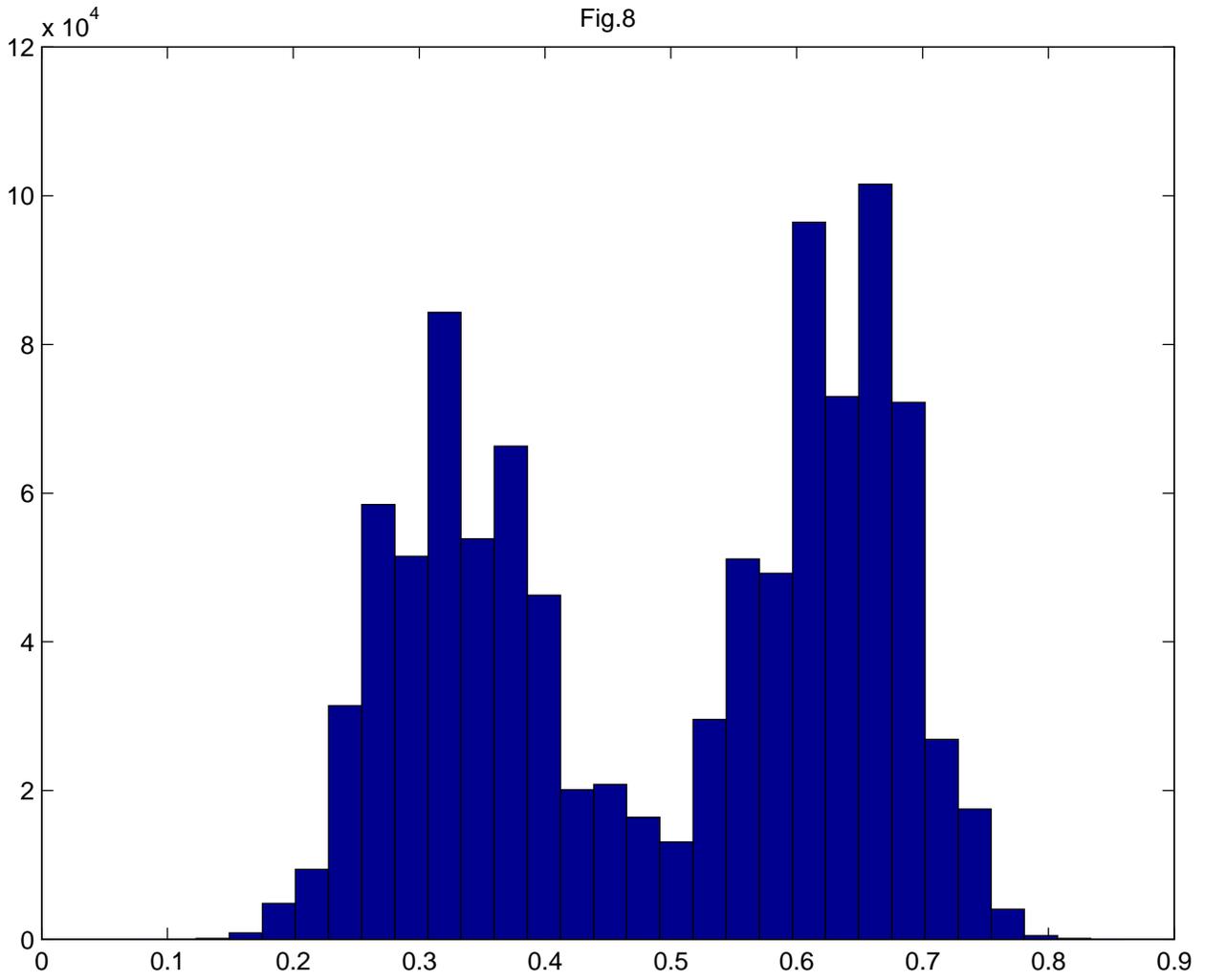
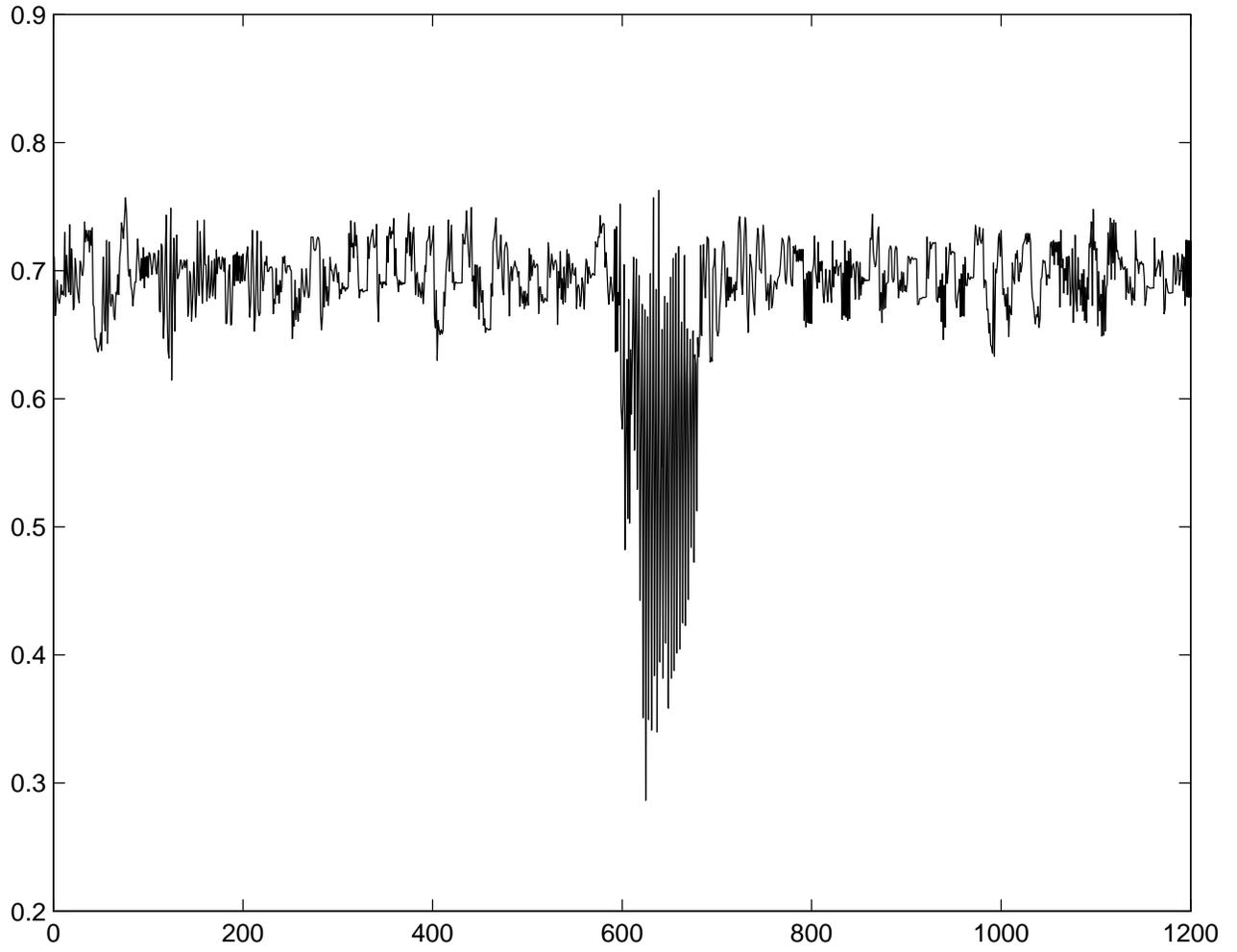
\end{document}